# High-speed Laser Micromachining with Copper Bromide Laser

Ivaylo I. Balchev[a,*], Nikolai I. Minkovski[a], Ivan K. Kostadinov[b], Nikola V. Sabotinov[a]

[a]Laboratory of Metal Vapor Lasers, Institute of Solid State Physics,
Blvd.Tzarigradsko Chaussee 72, 1784 Sofia, Bulgaria,
[b] Pulslight Co., Blvd.Tzarigradsko Chaussee 72, 1784 Sofia, Bulgaria

[*]Corresponding author. Tel/Fax: +359-2-9743002, e-mail: balchev@issp.bas.bg

## ABSTRACT

The application of the copper bromide (CuBr) laser as an attractive tool in the micro-machining of different materials has been demonstrated. High-quality drilling by trepanning and precision cutting was established on several materials with a negligible heat-affected zone (HAZ). That good performance was a result of the combination of high power visible radiation, short pulses, and close to the diffraction-limited laser beam quality with high-speed galvo scanner beam steering.

**Keywords:** CuBr laser, laser drilling, laser cutting, laser marking

## 1. INTRODUCTION

Laser machining can remove material in very small portions while traditional machining removes material can not. Laser machining processes are said to remove material by atomic layers. For this reason, the kerf in laser cutting is usually very narrow, the depth of laser drilling can be controlled to less than one micron per laser pulse and shallow permanent marks can be made with great flexibility (under flexibility, hereafter we mean generally the shape diversity of laser machining produce). In this way, material can be saved, which may be important for precious materials or for delicate structures in micro-fabrications. That also means small removal rate of laser machining compared to traditional machining. However, laser cutting of sheet material with thickness less than 2 mm can be fast, flexible and of high quality.

High power laser are a widely used in industrial manufacturing applications such as drilling, cutting, welding and surface processing where typically CO2 and Nd-YAG systems operating in the infrared are used. However, the laser material interactions in many machining cases are more effectively when using lasers operating at shorter wavelengths. For example, micro-machining of many metals is best performed using visible wavelengths. Copper vapor lasers have a good place in this area due to visible wavelength, high pulse repetition rate, high peak and average power, good beam quality, short pulse length [1]. When applied to materials processing, the visible wavelength is found to couple well to most materials, the short pulse length causes ablative material removal, and the high repetition rate increase the material removal rate [2]. Another advantage of the copper vapor laser is near diffraction-limited operation allowing precise spatial control.

In this paper we present our results in the micro-machining applications of our laser system based on CuBr master oscillator and power amplifier connected with computer controlled high-speed scanner system.

## 2. EXPERIMENTAL SETUP (laser drilling system)

Our laser system is a MOPA (Master Oscillator - Power Amplifier) CuBr laser with maximum output power 10W, divergence 120 μrad, repetition rate 19 kHz, and output beam diameter 15 mm, generating two wavelengths 511 nm and 578 nm in ratio 2:1. The master oscillator was formed by a discharge tube placed in an unstable confocal negative-branch resonator with magnification M=60 and perpendicular optical output through a plane mirror 45-degree tilted, with a small hole (0,5 mm) at the center. The power amplifier was a single pass with the same diameter of the active medium as the oscillator – 15 mm.

The temporal pulse shape was measured with a fast pin-photo diode (1 ns rise time) and Tektronix (200 MHz) oscilloscope. A typical form of the green (511 nm) laser pulse is present on the figure 1 (the pulse duration is approximately 30 ns). For the laser short-term stability, the pulse to pulse fluctuation in intensity and duration was found to be ±5%.

The master timing system (MTS) is a computer-controlled tool for synchronizing the laser oscillator and power amplifier [3]. It functions as an optical shutter by the controlling of the delay, too. The MTS provides the MOPA power supplies with triggering signals of controlled delay time. Depending on the delay the second laser acts as an amplifier or as an absorber (shutter). We have to note that best results for laser divergence we had at certain delay times. The optimum delay we found experimentally by varying the delay and measuring the divergence simultaneously. Usually in this case the output energy of the MOPA system was below the maximum but for micromachining applications is more important to have higher beam quality.

For micromachining applications very important is to have good stability of the beam shape and spot position after focusing. To observe the beam profile in the focus and to calculate the divergence we used Spiricon LBA-300 beam analyzer. Part of the output beam was send to a CCD camera focused by an achromatic lens of 1m-focal length. On the figure 2 is presented a typical beam distribution in the focus. The full angle divergence is 120 µrad. That is 1,5 times larger than the diffraction-limited divergence, which is 80 µrad for 511nm. The CCD camera of the beam analyzer had also an electronically shutter with minimum gate of 0.1ms, so this system is a good tool for measuring the pulse-to-pulse stability in the focus. In our case the fluctuation of the focal spot on the focal plane (point-to-point stability) was less than 10%.

The experimental set-up is shown in figure 3. Our high-precision machining experiment was based on this MOPA laser system and X-Y galvo scanner. The alignment in Z direction was done manually to find the best position of the focus via micro translation screw.

The laser beam was focused perpendicularly onto the sample surface by an objective (model F-Theta of focus of 32 cm) that guaranteed the place of the focal spot on the plane surface with dimension 12x12cm. This focusing system was not achromatic and only one laser line was used - the green (511nm) line. The yellow line was removed with ? dichroic mirror.

For the controlling the laser treatment process we used a CCD camera (EHD CamPro 02IR). The camera was coupled to an Optem objective with a zoom of 100. This camera was used to observe the forms and the quality of the cutting or drilling edges. Also, for a detailed investigation of the micro machined sample we used sometimes a scanning electronic microscope (SEM).

## 3. LASER DRILLING AND CUTTING

Laser drilling can be done by two ways: by trepanning and by percussion drilling (punching). In the punching process the laser beam is normally kept still and the hole is punched through the material using multiple pulses at a high repetition rate. The size and shape of the holes are governed by the size and shape of the focused laser beam (the dimension of the holes is usually the same as the diameter of the beam spot in the focus). The trepanning requires a relative beam-target movement during the processing. The parameters for trepanning are the same as for the cutting. There is a limit to the depth of material that can be cut in a single pass, so the number of passes has to be performed. The size and shape of holes can be programmed too. Holes formed by this method can be placed at very fine pitch and have a high aspect ratio (i.e. depth vs. width) required for multi-layered small geometry circuits. In this case, the rotations of the beam spot (or the target) allow making the process as well as the size and shape of holes programmable.

Laser punching and laser trepanning can produce through-holes. The choice comes from the hole size we need. Laser punching makes holes of diameter normally less than 50 μm. The laser trepanning holes are larger. Laser trepanning is a method by which the laser beam cuts in a circular pattern, taking advantage of high-speed beam positional scanner. There is a limit to the depth of material that can be cut in a single pass, so a number of passes has to be made. The number of the repetitions is defined as number of circles around the contour of a hole.

We applied the trepanning method for the better flexibility and hole quality. Another reason for choosing this method is our experimental set-up that allows fast modification of the processing conditions. We can control precisely the speed of trepanning, the number of repetitions, the acceleration of the scanner (this parameter is important for cutting figures of more complex pattern).

The targets in our experiments were copper and aluminum foils as materials of importance in electronics (PCB, etc.). Our goal was to find the best conditions for drilling holes of good circular shape, of minimum heat-affected zone (HAZ) and of minimum processing time. Laser processing of these materials has serious inherent problems for the high reflectivity and very high thermal conductivity of copper and aluminum. To minimize thermal material degradation outside the illuminated zone a limitation of heat transport is shown with the use of femtosecond, picosecond or nanosecond pulses. The HAZ depends on thermal diffusion length $L_D$ defined by $L_D = (\chi \tau_p)^{1/2}$ where $\tau_p$ is the laser pulse width and $\chi$ is the thermal diffusivity constant [4].

The experimental arrangement is shown in figure 3. Test samples were placed in a holder with a z-screw alignment. To measure breakthrough of the laser beam through the target foil a photodiode was placed under it. The photodiode was connected to MTS and when the hole was drilled the signal from the photodiode turned off the laser beam. During the drilling, the software of the scanner measured the number of repetitions and the real duration of the laser machining process.

The average laser power was measured by a laser power meter (Scientech Vector H310 model). Then we calculated the peak power of laser pulse.

Some preliminary experiments were performed with very encouraging results. A series of drilling trials were carried out at different velocity of the laser beam scanner and different average power of the laser. The copper and aluminum samples were with thickness 100 and 150 µm respectively. Before processing they were immersed in ethanol to remove surface contaminations as grease or oils. After the drilling tests had been carried out, the quality of the holes was investigated with a CCD camera. For a detailed examination of the holes we used scanning electronic microscope (SEM). Typical holes in aluminum and copper are presented in figures 4 and 5.

For these experiments we kept the laser power constant – 4 W and this correspond to a peak power in the focus of about 0,65 GW/cm$^2$ and pulse energy of 0,2 mJ. This power is not enough for one shot-drilling (or cutting) and the trepanning method ensures multi-pulse interactions.

We found that the trepanning reduces significantly the heat-affected zone when the motion velocity of the laser beam spot is relatively high. The velocity of the beam determines the degree of the overlapping of the laser pulses (the repetition rate of the laser pulses is kept constant). By overlapping, the leading edge of the second pulse meets surface temperature higher than the first pulse. So, we have an accumulation temperature effect that conduces to the sample surface high temperature [5]. We found out that the maximum efficiency depends on the extent of the laser pulse overlapping (i.e. scanner velocity).

On other side the target temperature increase is important because the reflectivity of the metals normally decreases as the surface temperature goes up, and the overall efficiency of the laser processing increases too [6]. This case is presented on the figure 4, where the drilling of Al foil was with good efficiency and negligible HAZ. Without overlapping, when the velocity is too high (in our case, more than 800mm/s), the HAZ is negligible but the efficiency of the drilling decreases sharply (we had to increase greatly the number of repetition cycles).

When the degree of overlapping is higher (lower velocities) the heat-affected zone expands considerably. Of coarse, in this case we observed a decrease of the number of repetitions but the integral time for drilling was longer (fig.5 - Cu). In this case the quality of the holes was worse, the HAZ increased and the edges of the holes were with an unwanted rim because part of the melted material resolidified on the edges. Around the hole were many ejected metal droplets and their amount increased with decreasing the scanner velocity.

We have to note that the hole walls drilled in copper and aluminum foils are very clear without any thermal hazards (fig.6).

Laser cutting is one of the most frequently used machining processes employing lasers. A typical laser for cutting is CW Nd-YAG or $CO_2$. Our experiments with CuBr laser as a cutting tool were also promising but only for metal foils. Although the quantity of removed material per pulse is small for these processes, the high repetition rate of CuBr laser allows material processing at high speeds. In addition, by using small laser pulse energies (typically 0,2-0,4 mJ), we can control the cut depth very precisely. The small size of the focused beam spot and high intensity allow us to achieve material cuts of very good quality and narrow cutting line. Cut widths of about 30 µm were obtained in copper and aluminum with good quality of the walls.

Cutting complex figures from these materials were also very successful. The edges of the figures were usually clear and sharp and there was no evident thermal distortion. However, problems appeared in areas around the corners. They are likely to be eliminated by increasing the extent of laser pulses overlapping there. This is the reason to get the surface temperature higher, thus increasing the HAZ and melting ejected metal droplets. A typical picture of $90^0$- corner cutting in copper foil is presented on the figure 7.

## 4. CONCLUSION

MOPA CuBr vapor laser system, combined with computer controlled galvo scanner was used to investigate the various machining processes for several materials.

The experimental results showed that our system is a good tool for high-speed laser drilling and cutting of wide variety of materials. In spite of the short period in which our CuBr laser system has been used for precision machining a number of promising results have been demonstrated. The trepanning method for drilling or cutting with kilohertz laser pulses allows better control of high-speed machining with good quality. The combination of computer controlled galvo scanner and high-quality and high–power laser pulses make laser processing very fast, precise for aluminum and copper.

## 5. ACKNOWLEDGMENTS

## 6. REFERENCES

[1] N.V. Sabotinov 1996 Pulsed Metal Vapour Lasers. NATO ASI Series 1/5, Kluwer Academic Press. 113-124

[2] J.S. Lash and R.M. Gilgenbah. 1993 *Rev. Sci. Instrum.* **64** 3308-13

[3] H.W. Bergmann, C.Korner, M.Hartmann, R.Mayerhofer. 1996 *Pulsed Metal Vapour Lasers.* NATO ASI Series 1/5, Kluwer Academic Press. 317-330

[4] M. El-Bandrawy, K. Nagarathnam, M. Gupta, C. Hamann and J. Horsting. (2003) *J. of Laser Appl.* **15** 101

[5] Hong Lei, Li Lijun. *Optics & Laser Technology* **31** 531

[6] G. Andra, E. Glauche. 1997 *Appl. Surf. Sci.* **109/110** 133

List of figure captions:

**Figure 1**. Typical temporal laser pulse shape generated from our MOPA system (the green laser line 510 nm) using 1 ns rise time photodiode.

**Figure 2**. Laser intensity distribution in the focal plane. The laser beam was focused with 1 m focal length achromatic lens.

**Figure 3**. Schematic of the experimental complex for laser micromachining.

**Figure 4**. Scanning electron micrograph showing the surface of 0,1 mm thickness Aluminum, drilled with our CuBr laser by trepanning process (laser power - 4 W, ?= 511nm, scanner velocity – 20 mm/s, number of repetitions – 30).

**Figure 5**. Scanning electron micrograph showing the surface of 0,15 mm thickness Copper, drilled with our CuBr laser by trepanning process (laser power - 4 W, ?= 511nm, scanner velocity – 2 mm/s, number of repetitions – 10).

**Figure 6**. Typical SEM view of a laser cutting wall of the hole (same in the fig.5 with relative magnification 15 times) in copper foil with thickness of 150 μm,

**Figure 7**. Scanning electron micrograph showing the picture of $90^0$- corner cutting in 0,1 mm thickness copper foil (laser power – 4 W, scanner velocity – 30 mm/s, number of repetitions - 20).

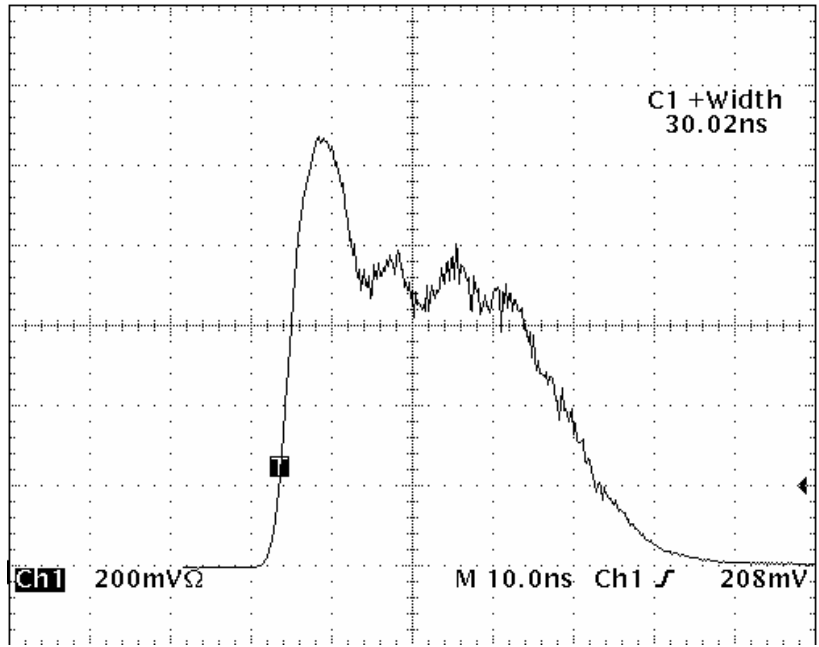

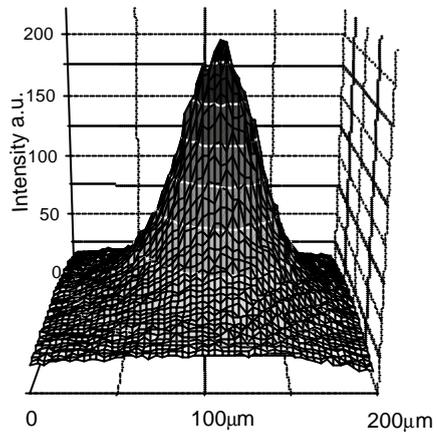

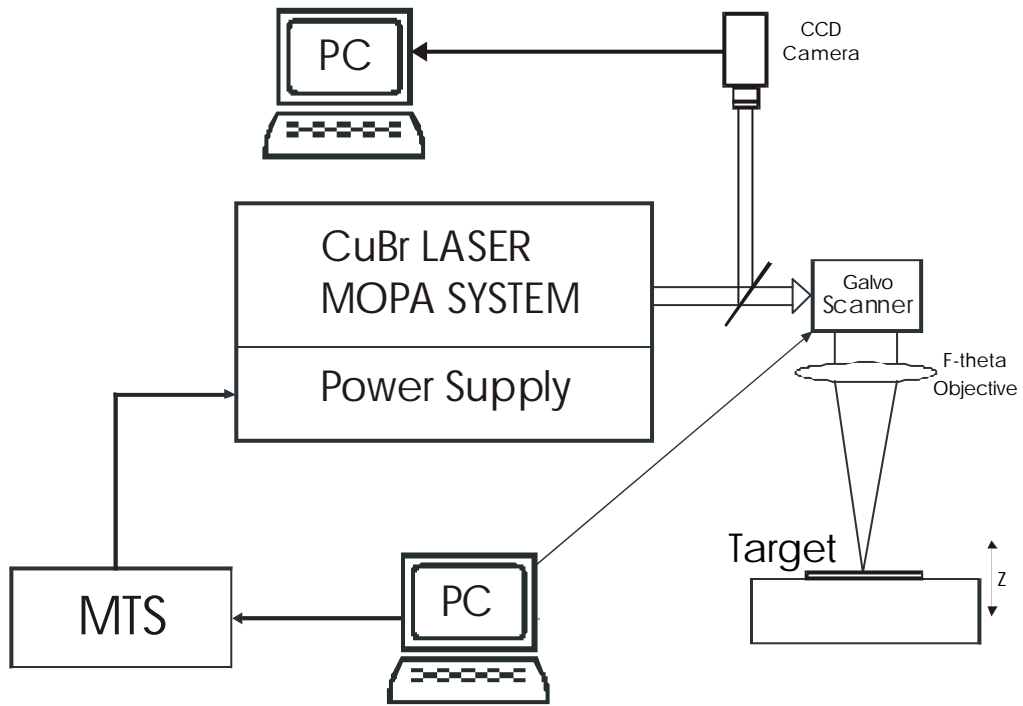

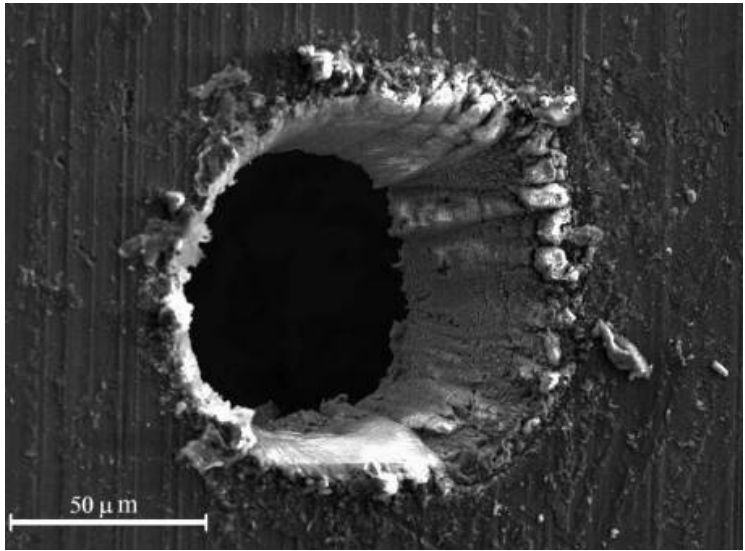

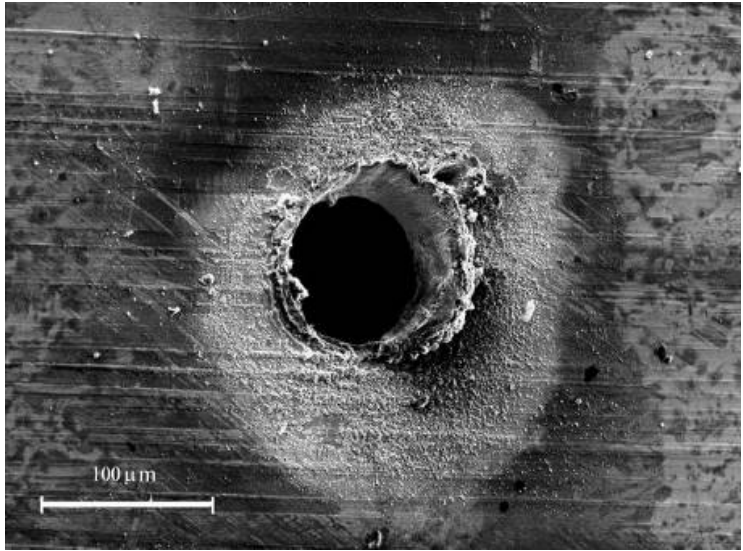

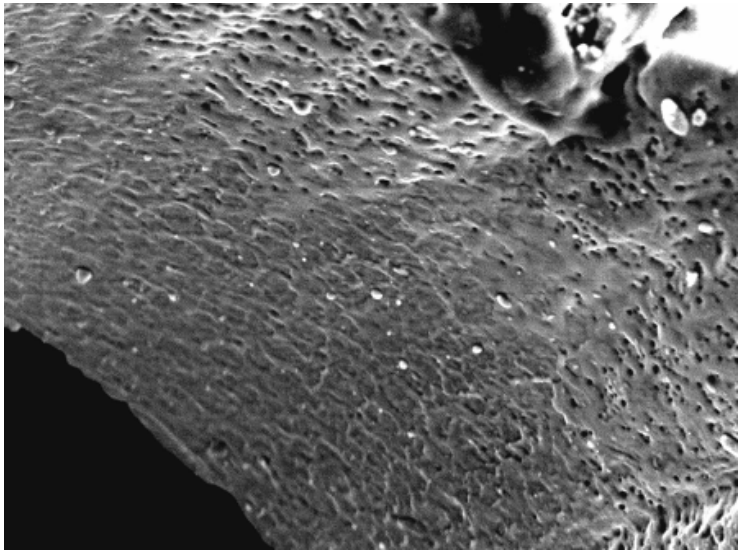

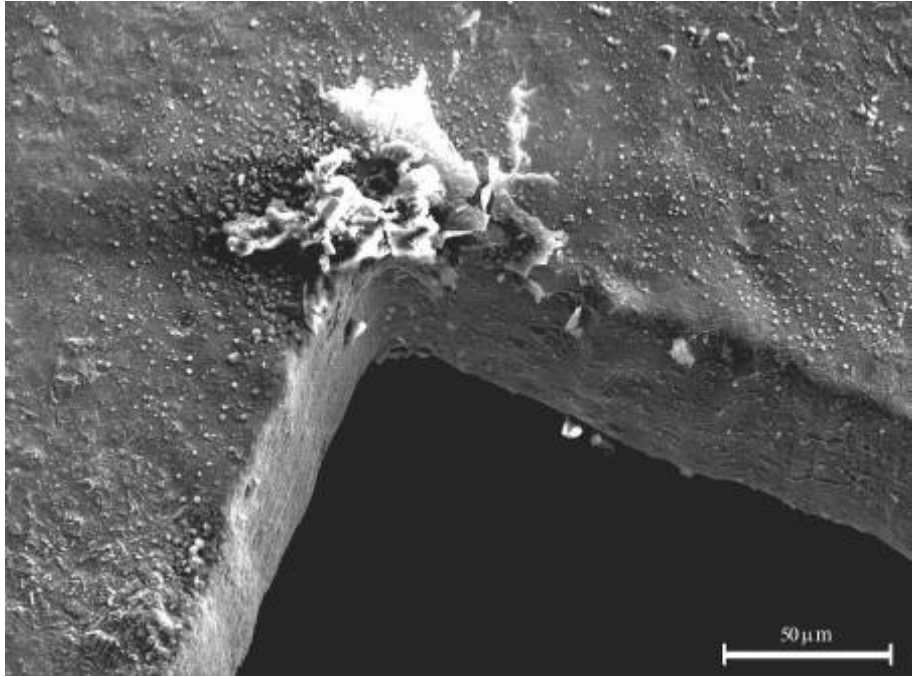